\documentclass[letterpaper,12pt]{article}

\renewcommand{\theequation}{\arabic{equation}}
\newcommand{\3}{\,}
\newcommand{\rd}{{\rm{d}}}

\newlength{\xa}
\newlength{\parin}
\begin{document}

\begin{center}
         {\large\bf{PLASMA DYNAMICS AND A SIGNIFICANT ERROR
         \vspace{1ex} \\ OF MACROSCOPIC AVERAGING}}
         \vspace{4ex}\\
         {\large{\bf{Marek A. Sza{\l}ek}}}
         \vspace{2ex}\\
         {\em{Institute of Fundamental Technological Research of the
         Polish Academy of Sciences \\
         \'{S}wi\c{e}tokrzyska 21, 00-049 Warsaw, Poland \\
         Address for correspondence: Krasnobrodzka 13 m. 94, 03-214
         Warsaw, Poland\\
         E-mail:$\;$ mszalek@poczta.tp.pl$\;\;$ }}
         \vspace{3ex}\\
\end{center}

            {\bf{Abstract:}} The methods of macroscopic averaging used to
         derive the macroscopic Maxwell equations from electron theory
         are methodologically incorrect and lead in some cases to a
         substantial error. For instance, these methods do not take into
         account the existence of a macroscopic electromagnetic field
         ${\bf E}_B$, ${\bf H}_B$ generated by carriers of electric
         charge moving in a thin layer adjacent to the boundary of the
         physical region containing these carriers. If this boundary is
         impenetrable for charged particles, then in its immediate
         vicinity all carriers are accelerated towards the inside of the
         region. The existence of the privileged direction of
         acceleration results in the generation of the macroscopic field
         ${\bf E}_B$, ${\bf H}_B$.  The contributions to this field from
         individual accelerated particles are described with a sufficient
         accuracy by the Li\'{e}nard-Wiechert formulas. In some cases the
         intensity of the field ${\bf E}_B$, ${\bf H}_B$ is significant
         not only for deuteron plasma prepared for a controlled
         thermonuclear fusion reaction but also for electron plasma in
         conductors at room temperatures. The corrected procedures of
         macroscopic averaging will induce some changes in the present
         form of plasma dynamics equations. The modified equations will
         help to design improved systems of plasma confinement.

\vspace{2ex}
\vspace{2ex}
\vspace{0ex}\\

\begin{center}
         {\bf{1. INTRODUCTION}}\nopagebreak
\end{center}
\setcounter{equation}{0}
\setlength{\parindent}{1\parin}
\vspace{-10\xa}

            The methods of macroscopic averaging used to derive the
         macroscopic Maxwell equations are based on two false
         mathematical assumptions and lead sometimes to false results.
         This will be explained in the present section.

            It has been directly shown in my earlier paper (Sza{\l}ek
         1997) that the correct procedure of macroscopic averaging of the
         electromagnetic field generated by particles from a small
         macroscopic region $P$ produces a field not always satisfying
         the macroscopic Maxwell equations. This follows from a simple
         example discussed in that paper or from comment 2 on p. 100
         there. This result may be considered a consequence of the fact
         that each contribution to the field from an individual carrier
         leaving or entering $P$ is discontinuous in time. It is
         different from zero when the generating carrier is in $P$ (in
         the retarded time) and is equal to zero otherwise.  Therefore,
         none of these contributions satisfies the pertinent microscopic
         Maxwell equations, which invalidates, e.g., the formulas (45),
         (46) on p. 67 and (84), (85) on p. 113 in the book of de Groot
         (1969) as well as formulas (8) on p. 24 and (62) on p. 256 in
         the book of de Groot and Suttorp (1972). These formulas average
         the microscopic Maxwell equations with pseudo-solutions
         introduced to these equations, where the pseudo-solutions are
         the contributions to the electromagnetic field from individual
         carriers. The aim was to obtain the macroscopic electromagnetic
         field being the solution of the macroscopic Maxwell equations.
         This procedure is erroneous as the pseudo-solutions do not
         satisfy the pertinent microscopic Maxwell equations if the
         boundary of $P$ is penetrable for carriers.

            If a proof is erroneous, it need not necessarily mean that
         its thesis is incorrect. That, however, is not the case here,
         and the proper solutions of the macroscopic Maxwell equations
         not always describe correctly the macroscopic electromagnetic
         field. The differences between these solutions and the real
         macroscopic fields are sometimes surprisingly great. The proof
         is given in this paper in some cases when the macroscopic
         electric current does not flow through the boundary of the
         region $P$. Such an outcome has been already suggested by
         results of my earlier paper concerning some cases when the
         macroscopic current flows through the boundary of $P$. The
         results obtained in the present paper make it possible to
         propose in Section 2 a convenient procedure which will allow to
         calculate the correct value of the averaged electromagnetic
         field acting on charged particles in material media.

            The derivation of the macroscopic Maxwell equations is also
         accompanied by another methodological error. Namely, the
         pertinent considerations take into account the distribution
         functions of particles depending only on space coordinates
         and on the velocities of carriers of electric charge. They
         disregard the distribution of carrier accelerations.
         Consequently, the particles with the same velocities but
         different accelerations are treated identically, although the
         fields generated by these particles are different. Such an
         approach may be justified when in a small macroscopic region, or
         in a thin layer adjacent to a privileged surface, the average
         acceleration of all carriers is equal to zero. This approach is
         incorrect when we consider a thin layer adjacent to the
         impenetrable boundary of a physical medium, where the average
         acceleration is different from zero and is directed towards the
         inside of the medium. It is also erroneous when we take into
         account an macroscopically inhomogeneous medium.

            The considerations of this paper are based on the existence
         of the macroscopic electromagnetic field generated by
         carriers of electric charge accelerated in the immediate
         vicinity of impenetrable boundaries. This field has not been
         taken into account in the pertinent argumentation of other
         authors. Its intensity is sometimes astonishingly great, which
         indicates that the whole problem of the macroscopic
         electromagnetic field in material media should be meticulously
         reconsidered.

            This paper is an abbreviated and simplified version of my
         earlier paper "Plasma dynamics and boundary phenomena in
         macroscopic averaging" formulated in 2002/2003. In that paper
         results concerning the correct calculation of the macroscopic
         electromagnetic field have been obtained in an exact and
         systematic manner in the general case of curved boundaries and
         real physical media. Certain considerations there, however, seem
         rather difficult for some physicists, and the exactness of the
         argumentation has made that paper relatively long (25 pages). To
         make the argumentation of the present paper shorter and simpler,
         we idealize to some extent the properties of the media in the
         vicinity of the impenetrable boundary. We also assume that the
         boundaries are plane. Nevertheless, the results obtained here
         agree with the results of that earlier paper. These of the more
         important results of the latter which have not been derived
         here, are repeated without proofs in Section 5. We have also
         retained some methodological remarks which are given in Section
         6. That earlier paper is available from the author on request
         electronically in a form of a pdf file.
\vspace{0ex}\\

\begin{center}
         {\bf{2. INITIAL CONSIDERATIONS}}\nopagebreak
\end{center}
\setcounter{equation}{0}
\vspace{-10\xa}

            Consider a macroscopic bounded region $P$ containing a plasma
         or a typical metal without significant magnetic properties. We
         denote the boundary of $P$ by $\partial P$ and assume that
         $\partial P$ is composed of two sufficiently regular surfaces
         $\partial_1 P$ and $\partial_2 P$. The surface $\partial_2 P$ is
         impenetrable for carriers of electric charge.

            Let ${\bf J}$ and $\rho$ be the macroscopic densities of,
         respectively, electric current and charge in $P$, and let the
         normal to $\partial{P}$ component of ${\bf J}$ vanish
         identically on $\partial{P}$. Let ${\bf E}_0$, ${\bf H}_0$
         denote respectively the electric and magnetic field being the
         solution of the Maxwell equations with the densities ${\bf J}$
         and $\rho$, and satisfying the proper boundary conditions.  If
         ${\bf J}$$\,\equiv\,$$0$, $\rho $$\,\equiv\,$$ 0$ in $P$, then
         ${\bf E}_0 $$\,\equiv\,$$ 0$, ${\bf H}_0 $$\,\equiv\,$$ 0$.
         According to the present opinions the field ${\bf E}_0$, ${\bf
         H}_0$ is equal to the macroscopic electromagnetic field $\bf E$,
         $\bf H$ generated by particles from the region $P$:
\begin{eqnarray}
         {\bf E} = {\bf E}_0, & & {\bf H} = {\bf H}_0.   \label{1.a}
\end{eqnarray}
            The field $\bf E$, $\bf H$ is the sum of contributions to the
         macroscopic electromagnetic field from all particles in $P$. We
         shall show that in order to describe correctly this field, one
         should take into account in some cases not only the field ${\bf
         E}_0$, ${\bf H}_0$, but also two other macroscopic fields which
         will be denoted by ${\bf E}_B$, ${\bf H}_B$ and ${\bf E}_T$,
         ${\bf H}_T$. The field ${\bf E}_B$, ${\bf H}_B$ is generated by
         carriers accelerated in the immediate vicinity of the boundary
         $\partial_2P$. For all carriers this acceleration is directed
         towards the interior of $P$ and is induced by forces which make
         $\partial_2P$ impenetrable. The existence of the privileged
         direction of acceleration causes that the sum of contributions
         to the field ${\bf E}_B$, ${\bf H}_B$ from particular carriers
         does not vanish. It will be shown that the intensity of the
         field ${\bf E}_B$, ${\bf H}_B$ can be significant even in some
         cases of electron plasma in conductors at room temperatures. We
         shall refer to this field as ``the boundary acceleration
         electromagnetic field.''

            If ${\bf E}_0 $$\,\equiv\,$$ 0$, ${\bf H}_0$$\,\equiv\,$$ 0$,
         then the field ${\bf E}_T$, ${\bf H}_T$ is the sum of
         contributions, respectively, to the macroscopic electric and
         magnetic field, generated by all particles from $P$ with the
         exception of the contributions used to calculate ${\bf E}_B$,
         ${\bf H}_B$. If ${\bf E}_0 $$\,\not=\,$$ 0$ or ${\bf
         H}_0$$\,\not=\,$$ 0$, then we define that this sum is equal to
         ${\bf E}_0+{\bf E}_T$, ${\bf H}_0+{\bf H}_T$ We shall call the
         field ${\bf E}_T$, ${\bf H}_T$ ``the temperature electromagnetic
         field.''

            Let $[{\bf E,H}]$ be an ordered pair and $[{\bf
         E,H}]$$\,=\,$$[{\bf e,h}]$ if and only if ${\bf E}$$\,=\,$${\bf
         e}$ and ${\bf H}$$\,=\,$${\bf h}$. Let
\begin{equation}
         {\bf G}=[{\bf E},{\bf H}],\,\,\,\,\,\,\,\,
         {\bf G}_0=[{\bf E}_0,{\bf H}_0],\,\,\,\,\,\,\,\,
         {\bf G}_B=[{\bf E}_B,{\bf H}_B],\,\,\,\,\,\,\,\,
         {\bf G}_T=[{\bf E}_T,{\bf H}_T]. \,\,\,\,\,\,\,\,
    \label{1.1a}
\end{equation}
         Using this notation we may write the correct expression for the
         field ${\bf G}$ in the form
\begin{eqnarray}
         & {\bf G} = {\bf G}_0 + {\bf G}_B + {\bf G}_T.
         &  \label{1.1}
\end{eqnarray}
            This and the next equations concern the fields ${\bf{G}}$,
         ${\bf{G}}_B$, ${\bf{G}}_T$ in observation points lying outside
         the region $P$ and its immediate vicinity.  The fields
         ${\bf{G}}_B$ and ${\bf{G}}_T$ may be ignored when
\begin{eqnarray}
         & {\bf G}_B + {\bf G}_T = 0. &  \label{1.2}
\end{eqnarray}
         We shall show, however, that in some cases equation
         (\ref{1.2}) is not satisfied.

            First, consider the case when the plasma in $P$ is
         homogeneous and isotropic. Let $|{\bf G}|=|{\bf E}|+|{\bf H}|$
         where $|{\bf E}|$, $|{\bf H}|$ are the lengths of ${\bf E}$ and
         ${\bf H}$. It follows from the results obtained in the present
         paper that if $\partial_2P=\partial P$, then $|{\bf G}_B|$
         vanishes or is relatively small. We may expect the same from
         $|{\bf G}_T|$ taking into account equation (\ref{1.2}), the
         validity of which is in this case suggested, e.g., by the
         experimental fact that the electromagnetic field produced by
         carriers from any region $P$ with impenetrable boundaries, and
         averaged in a short interval of time, is at the outside of $P$
         and its immediate vicinity equal to zero with a good accuracy
         for media under consideration if the macroscopic densities ${\bf
         J}$ and $\rho$ vanish in $P$.

            The situation changes substantially when $\partial_2P \neq
         \partial P$ and $\partial_2P$ is not a closed surface. In such
         cases $|{\bf G}_T|$ remains roughly the same while $|{\bf G}_B|$
         can become surprisingly large. For instance, one impenetrable
         face of a copper cube with the length of edge equal to 1{\3}{cm}
         produces at a temperature of 300\3{K} the fields ${\bf E}_B$,
         ${\bf H}_B$ which in some points at a distance of 1{\3}m from
         the center of the cube have the intensity of about,
         respectively, $10^{11}\rm\3{V}\3{m}^{-1}$ and
         $10^8\rm\3{A}\3{m}^{-1}$. If the cube were filled with a
         rarefied deuteron plasma prepared for a controlled thermonuclear
         fusion reaction, then the corresponding intensities would be
         about $10^6\rm\3{V}\3{m}^{-1}$ and $10^4\rm\3{A}\3{m}^{-1}$. We
         can obtain then from~(\ref{1.1})
\begin{eqnarray}
         & |{\bf G}-{\bf G}_0| = |{\bf G}_B+{\bf G}_T| \gg 0. &
         \label{1.3}
\end{eqnarray}
            In particular, we can get $|{\bf G}| \gg 0$ when ${\bf
         J}$$\,\equiv\,$$ 0$, $\rho$$\,\equiv\,$$ 0$ in $P$. A proof of
         inequality (\ref{1.3}) for ${\bf G}_0\,$$\equiv$$\,0$ is given
         in Section 3.

            Inequality (\ref{1.3}) is interesting from the theoretical
         point of view, because it evidently disagrees with equations
         (\ref{1.a}) and with some universally accepted opinions
         concerning the macroscopic electromagnetic field generated by
         large sets of charged particles in chaotic motion. There are two
         main reasons of the disagreement between the results obtained in
         the present paper and the pertinent results of other authors.
         First, the methods of macroscopic averaging of electromagnetic
         fields applied in the scientific literature do not take into
         account the existence of the macroscopic boundary acceleration
         field ${\bf G}_B$ generated in a locally microscopic layer
         adjacent to $\partial_2P$. Second, it has not been noticed that
         for inhomogeneous or anisotropic media the temperature field
         ${\bf G}_T$ may be different from zero and attain considerable
         intensities. An additional reason of the disagreement is the
         discontinuity in time of the contributions to the macroscopic
         electromagnetic field from individual carriers, which has been
         already explained in Section 1.

            If the medium in $P$ is inhomogeneous and/or anisotropic,
         then $|{\bf G}_B|$ can become relatively very large even if
         $\partial_2P=\partial P$ or $\partial_2P$ is a closed surface.
         This may have practical consequences. It will be explained in
         the following example.

            Suppose that we want to calculate an averaged electromagnetic
         field ${\bf G}'=[{\bf E}',{\bf H}']$ acting on electrons and ions at a
         point $O$ of a region $P'$ with a boundary $\partial P'$. There
         are various methods of calculating this field. All of them need
         some corrections because they do not allow for the existence of
         the fields ${\bf G}_B$, ${\bf G}_T$. We shall point out a method
         where the manner of correction is obvious.

            Let the medium in $P'$ be inhomogeneous and the surface
         $\partial P'$ be impenetrable for carriers. To facilitate the
         considerations we assume that the medium inhomogeneities are
         stable, as in inhomogeneous alloys of metals. Consider in $P'$ a
         small subregion $P_S$ with a boundary $\partial{P_S}$ such that
         $O\!\,\in\!\, {P_S}$ and lies possibly far from $\partial
         {P_S}$. We may write down
\begin{equation}
         {\bf G}' = {\bf G}_S + {\bf G}    \label{1.4}
\end{equation}
         where the field ${\bf G}_S$ is generated by particles from
         inside the region $P_S$, and the field ${\bf G}$ is produced by
         particles from inside the region $P$ defined by $P=P'-P_S$. The
         field ${\bf G}_S$ may be calculated by means of summing the
         contributions from all particles in $P_S$ with the exception of
         a privileged particle, while the field ${\bf G}$ is calculated
         by means of the macroscopic densities ${\bf J}$ and $\rho$ in
         $P$. The standard practice is to assume that the latter field
         satisfies (\ref{1.a}), i.e. ${\bf G} = {\bf G}_0$.

            On the ground of results obtained in this paper it can be
         shown that the assumption (\ref{1.a}) is not always correct even
         if ${\bf G}_0=0$. This is a consequence of the facts that
         $\partial P= \partial P'+ \partial P_S$, and that the surface
         $\partial P_S$ is penetrable for carriers and does not generate
         a contribution to the field ${\bf G}_B$.  Let the contribution
         generated at $\partial P'$ to the field ${\bf G}_B$ be ${\bf
         G}'_B$ and let the normal to $\partial P_S$ component of ${\bf
         J}$ vanish on $\partial P_S$. We denote by ${\bf G}_{BS}$ the
         contribution to the boundary acceleration field which would be
         produced at $\partial P_S$ by particles from $P\/$ if $\partial
         P_S$ were impenetrable for carriers. In such a case the whole
         boundary $\partial{P}$ of $P$ would be impenetrable for
         carriers. Therefore, we obtain from (\ref{1.2})
\begin{equation}
         {\bf G}_T + {\bf G}'_B + {\bf G}_{BS} = 0.   \label{1.6}
\end{equation}
         We receive from (\ref{1.1}) when $\partial P_S$ is penetrable:
\begin{equation}
          {\bf G} = {\bf G}_0 + {\bf G}'_B + {\bf G}_T.   \label{1.7}
\end{equation}
         From (\ref{1.6}) and (\ref{1.7}) we get
\begin{equation}
         {\bf G} = {\bf G}_0 - {\bf G}_{BS}.   \label{1.8}
\end{equation}
         An example is given in Section 4, which indicates that in
         inhomogeneous plasma one can obtain $|{\bf G}_{BS}|\gg 0$ as in
         (\ref{5.3.10}). Consequently, instead of the incorrect formula
         (\ref{1.a}), we should use equation (\ref{1.8}). From
         (\ref{1.4}) and (\ref{1.8}) we get the correct expression for
         the field~${\bf{G}}'$:
\begin{equation}
       {\bf G}' = {\bf G}_S + {\bf G}_0 - {\bf G}_{BS}.   \label{1.9}
\end{equation}

            At this stage of the research one cannot exclude the
         possibility that the presence of the term ${\bf G}_{BS}$ in
         (\ref{1.9}) can have a noticeable influence on the value of
         ${\bf G}'$. Such an influence should cause some changes in the
         present form of plasma dynamics equations. One can expect the
         greatest values of $|{\bf G}_{BS}|$ in the regions where the
         strongest inhomogeneities are present, e.g., on a shock wave.

            In general, the field ${\bf G}_{BS}$ will depend on the
         derivatives of plasma density and temperature. Hence, its
         existence could cause some changes of values of the coefficients
         of partial derivatives in the plasma dynamics equations. In
         particular, some coefficients equal to zero in the present form
         of these equations might become different from zero. Even small
         differences may affect, e.g., results concerning the growth and
         propagation of plasma instabilities. In order to examine the
         consequences of equation (\ref{1.9}) one should derive the
         explicit dependence of the field ${\bf G}_{BS}$ on the
         macroscopic parameters of inhomogeneous plasma, and investigate
         relations between ${\bf G}_{BS}$ and the coefficients in the
         plasma dynamics equations.

            The proper methodology pointed out in this paper will be
         useful not only when one examines the consequences of equation
         (\ref{1.9}) but also when it is possible to investigate the
         influence of some additional forces generated by electrically
         charged particles. These forces have been discovered by the
         present author and are briefly described in an earlier paper
         (Sza{\l}ek 1997). Allowing for all forces produced by
         electrically charged particles, one will be able to obtain,
         e.g., an adequate description of phenomena occurring in boundary
         regions of globular lightnings. Such a description will be
         connected with further changes in the plasma dynamics equations.
         The modified equations will help to design improved systems of
         plasma confinement.

            The temperature and boundary acceleration fields are
         generated in various media and should be taken into account in
         the completed macroscopic theory of electromagnetic fields based
         on the microscopic properties of electrons, ions, and atoms.
         This theory came into existence a century ago, although some of
         its important premises were known much earlier -- compare, e.g.,
         the early theories of dielectrics or Amp\`{e}re's hypothesis
         that magnetism is due to micro-currents. First results
         concerning the derivation of macroscopic Maxwell equations from
         electron theory were obtained by Hendrik A. Lorentz (see, e.g.,
         Lorentz 1902, 1904, 1915). A list of authors working on this
         problem and a historical survey were given by de~Groot (1969).
         First results concerning the fields ${\bf G}_B$ and ${\bf G}_T$
         were published by the present author (Sza{\l}ek 1997). That
         paper contains, e.g., an estimate of the ratio $|{\bf H}_B+{\bf
         H}_T|/|{\bf H}_0|$ (using the nomenclature introduced here) in
         some cases when $P$ is a cylinder and $\partial_2P$ is the
         lateral surface of this cylinder. Because of lack of funds for
         the pertinent research, my next paper on this subject was
         written with a delay of more than 5 years.
\vspace{0\xa}\\
\begin{center}
         {\bf{3. A PROOF OF INEQUALITY (\ref{1.3}) }}\nopagebreak
\vspace{-5\xa}
\end{center}
\noindent
         {\bf{3.1. The field of a moving charge}}\nopagebreak
\vspace{5\xa}

            We shall calculate in sections 3.2 and 3.3 the contributions
         to the macroscopic electromagnetic field from particular
         carriers using some approximate expressions resulting from the
         exact Li\'{e}nard-Wiechert formulas (Li\'enard 1898, Wiechert
         1900). In the SI systems of units these expressions may be
         written down in the form
\begin{eqnarray}
    {\bf e(x},t)= {\bf e}_a+{\bf e}_u, & &
    {\bf h(x},t)={\bf h}_a+{\bf h}_u, \label{2.4.3}
\end{eqnarray}
         where ${\bf{e(x}},t)$, ${\bf{h(x}},t)$ are, respectively, the
         electric and magnetic field generated by a moving point charge
         $Q$ and observed at a point $\bf{x}$ at time $t$,
\begin{eqnarray}
         {\bf e}_a({\bf x},t) = L_1 R_1^{-1}{\bf{n}}_1
         {\bf\times} ({\bf{n}}_1{\bf\times}{\bf{a}}),\;\;\;\;\;\;\;\;\;\;
      &\;\;\;\;&
         {\bf h}_a({\bf x},t) = c\varepsilon_0 {\bf n}_1
         {\bf\times}{\bf e}_a, \;\;\;\;\;\;\;\;\;\;
       \label{2.4.5}\\
         {\bf e}_u({\bf x},t) = c^2L_1 (1-u^2/c^2)
         w^3R_1^{-2}({\bf{n}}_1-{\bf{u}}/c),
      &\;\;\;\;&
         {\bf h}_u({\bf x},t) = c\varepsilon_0 {\bf n}_1
         {\bf\times}{\bf e}_u, \;\;\;\;\;\;\;\;\;\;
       \label{2.4.4}
\end{eqnarray}
         where ${\bf R}_1(t)={\bf x-r}(t)$; $\;\;\bf{x}$, $\bf{r}$
         are, respectively, the position vectors of the observation point
         and the point charge in a Cartesian rectangular coordinate
         system $X$ with the coordinates $x_1$, $x_2$, $x_3$;
         $\;\;R_1=|{\bf R}_1|$, $\;\;{\bf{n}}_1(t)={\bf R}_1/R_1$;
         $\;\;\bf{a}$, $\bf{u}$ are, respectively, the acceleration and
         velocity of the point charge; $\;\;\varepsilon_0=
         8.85\!\times\!10^{-12}{\rm\3{F}\3{m}^{-1}}$, $\;\;c$ is the
         speed of light, and
\begin{eqnarray}
         L_1=Q(4\pi\varepsilon_0 c^2)^{-1},
   & &   w=(1-{\bf{n}}_1\cdot {\bf{u}}/c)^{-1}.
       \label{2.4.8}
\end{eqnarray}
         We have neglected in (\ref{2.4.5}) and (\ref{2.4.4}) the time
         retardation, and in (\ref{2.4.5}) we have retained only the
         first term of the expansion with respect to the powers of $u/c$.
         The exact formulas may be found in various books and monographs
         (see, e.g., Clemmow and Dougherty 1969). They are also used in
         my previous pertinent paper.
\vspace{0\xa}\\

\noindent
         {\bf{3.2. Introductory considerations}}\nopagebreak
\vspace{5\xa}

            Consider two identical macroscopic cubes $P^{(i)}$ ($i=1,2$)
         containing a suitable homogeneous medium with the macroscopic
         properties not depending on time. We assume for the sake of
         simplicity that the ions of the medium are at rest in $P^{(i)}$
         (see, however, remark~2 in Section~5). The cube $P^{(1)}$ is
         inside the medium and its boundary $\partial
         P^{(1)}=\partial_1P^{(1)}$ is fully penetrable for electrons.
         The cube $P^{(2)}$ is adjacent to the boundary of the medium and
         one of its faces (denoted by $\partial_2P^{(2)}$) is
         impenetrable for electrons while the remaining boundary
         $\partial_1P^{(2)}$ of $P^{(2)}$ is penetrable.

            Let ${\bf{G}}^{(i)}$, ${\bf{G}}_B^{(i)}$, ${\bf{G}}_T^{(i)}$
         denote, respectively, the macroscopic electromagnetic fields,
         the boundary acceleration electromagnetic fields, and the
         temperature electromagnetic fields generated by charged
         particles from $P^{(i)}$, and let ${\bf{G}}_0^{(i)}$ be the
         proper solutions of the macroscopic Maxwell equations in
         $P^{(i)}$ ($i=1,2$). These fields will be calculated in
         observation points lying outside the cubes $P^{(i)}$,
         sufficiently far from their boundaries. We shall prove that at
         least one of the fields ${\bf{G}}^{(i)}$ satisfies the formula
         (\ref{1.3})

            Assume that the macroscopic current and charge densities in
         $P^{(i)}$ are identically equal to zero. Therefore,
\begin{equation}
         {\bf{G}}_0^{(i)} \equiv 0, \;\;\;\;\;\;\;\;\;\;\;\;\;\;\;i=1,2.
         \label{m1}
\end{equation}
         Let ${\bf e}_a$, ${\bf h}_a$, ${\bf e}_u$, ${\bf h}_u$ be
         given by (\ref{2.4.5}) and (\ref{2.4.4}). The equation
         (\ref{m1}) being satisfied, we can calculate the field
         ${\bf{G}}_T^{(1)}$ by summing up the contributions [${\bf
         e}_a$,${\bf h}_a$], [${\bf e}_u$,${\bf h}_u$] from all
         particles in $P^{(1)}$ (i.e. from electron carriers and ions in
         $P^{(1)}$, the contributions from ions being given in our case
         by the Coulomb field). These contributions are averaged in a
         short interval of time when necessary.

            Assume now that the barrier of forces making the face
         $\partial_2 P^{(2)}$ impenetrable is sufficiently strong and
         steep to decelerate and accelerate the carriers within a
         distance $\kappa$ which is small in comparison with the average
         distance between carriers in $P^{(2)}$. Let $\Delta P^{(2)}$ be
         a boundary layer with a thickness $\kappa$ adjacent to the face
         $\partial_2 P^{(2)}$, and $\Delta P^{(1)}$ be a corresponding
         layer in $P^{(1)}$. To calculate ${\bf{G}}_T^{(2)}$, we sum up
         the contributions [${\bf e}_a$,${\bf h}_a$], [${\bf e}_u$,${\bf
         h}_u$] from all carriers and ions in $P^{(2)}-\Delta P^{(2)}$
         and add the contributions [${\bf e}_u$,${\bf h}_u$] from all
         particles in $\Delta P^{(2)}$. We do not take into account the
         contributions [${\bf e}_a$,${\bf h}_a$] from $\Delta P^{(2)}$,
         because they are used to calculate the field ${\bf{G}}_B^{(2)}$.

            If $\kappa$ is sufficiently small, then in the corresponding
         points of observation
\begin{equation}
         {\bf{G}}_T^{(1)}={\bf{G}}_T^{(2)}={\bf{G}}_T,       \label{m2}
\end{equation}
         since the properties and velocity distributions of particles
         are identical in $P^{(i)}-\Delta P^{(i)}$, and we may neglect
         the contributions [${\bf e}_a$,${\bf h}_a$], [${\bf e}_u$,${\bf
         h}_u$] from particles in $\Delta P^{(1)}$ and the contributions
         [${\bf e}_u$,${\bf h}_u$] from particles in $\Delta P^{(2)}$, as
         the bounds of all these contributions do not depend on~$\kappa$.
         The equation (\ref{m2}) being satisfied, we shall need no more
         information concerning the fields ${\bf{G}}_T^{(i)}$ to prove
         the formula (\ref{1.3}). However, we must calculate the fields
         ${\bf{G}}_B^{(i)}$. We have
\begin{equation}
                     {\bf{G}}_B^{(1)} \equiv 0,       \label{m3}
\end{equation}
         because all contributions [${\bf e}_a$,${\bf h}_a$], [${\bf
         {e}}_u$,${\bf {h}}_u$] from all particles in $P^{(1)}$ have been
         already used to calculate the field ${\bf{G}}_T^{(1)}$. It
         remains to calculate ${\bf{G}}_B^{(2)}$.
\vspace{0\xa}\\

\noindent
         {\bf{3.3. The field ${\bf{G}}_B^{(2)}$}}\nopagebreak
\vspace{10\xa}

            Let points ${\bf{x}}'$ with coordinates $x'_1$, $x'_2$,
         $x'_3$ in the Cartesian system $X$ denote points belonging to
         $P^{(2)}$. We define for a positive $\nu$ that $P^{(2)}$ is
         given by formulas: $\;x'_1,x'_2\in[-\nu,\nu]$,
         $\;x'_3\in[-2\nu,0]$, and that $\partial_2 P^{(2)}$ lies on the
         plane $x'_3=0$. Let $\bf{K}$ be a unit vector normal to the
         plane $x'_3=0$ and pointing in the positive direction of the
         $x_3$-axis.

            Assume for the sake of simplicity that all electron carriers
         move only in the $x_3$-direction, and that all of them have the
         same speed $u$ outside the boundary layer $\Delta P^{(2)}$ (the
         results obtained by means of this assumption may be easily
         generalized for some more general cases). In the layer $\Delta
         P^{(2)}$ each electron stays during a time $t_{\kappa}$ while it
         is accelerated from the velocity $+u{\bf{K}}$ to the velocity
         $-u{\bf{K}}$. Integrating ${\bf e}_a({\bf x},t)$ from
         (\ref{2.4.5}) with respect to time during the time $t_{\kappa}$,
         we obtain the averaged in a unit time contribution $\delta{\bf
         E}_B^{(2)}({\bf{x}})$ to the field ${\bf{E}}_B^{(2)}$ from one
         carrier. Let $\bf{R}={\bf{x}}$, $\:\: R=|\bf{R}|$,
         $\;\;{\bf{n}}={\bf{R}}/R\;\;$ ($R$ is the distance from the
         center of the face $\partial_2 P^{(2)}$ to the observation
         point). If $\;R\gg\nu+\kappa\;$ then with a good accuracy
         ${\bf{R}}_1\,$=$\:{\bf{R}}$ and $\;\delta{\bf
         {E}}_B^{(2)}({\bf{x}}) =-2L_1 u R^{-1}{\bf{n}\times}
         ({\bf{n\times{K}}})$.

            If the number density of electron carriers in $P^{(2)}$ is
         $\Phi_0$, then the number of collisions with the layer $\Delta
         P^{(2)}$ per unit area and time is $u\Phi_0/2$. Therefore, we
         obtain ${\bf{E}}_B^{(2)}({\bf{x}}) =-\Delta{S}\, L_1 \Phi_0 u^2
         R^{-1}{\bf{n}\times}({\bf{n\times{K}}})$ where $\Delta{S}$ is the
         area of $\partial_2 P^{(2)}\;$. Taking into account
         (\ref{2.4.8}), we can write
\begin{equation}
             {\bf{E}}_B^{(2)}({\bf{x}}) = -\Delta S\, L R^{-1}\,
         {\bf{n}} {\bf\times} ({\bf{n}} {\bf\times} {\bf{K}})
         \label{m4}
\end{equation}
         where
\begin{eqnarray}
         L = (2\pi\,m\,\varepsilon_0c^2)^{-1}\, Q\, \Phi_0\,W',
    & &  W'= mu^2/2,
         \label{m5}
\end{eqnarray}
         $m$ is the mass of electron.

            Similarly, integrating ${\bf h}_a({\bf x},t)$ from
         (\ref{2.4.5}), we get
\begin{equation}
         {\bf H}_B^{(2)}({\bf x}) = \Delta S\, c \varepsilon_0 L R^{-1}\,
         {\bf n} {\bf\times} {\bf K}.
         \label{m6}
\end{equation}
         Using the above method of derivation, one can also obtain the
         expressions for ${\bf G}_B^{(2)}$ in the general case when the
         electrons move in 3 space directions and have different
         velocities. In such a case, one gets the same expressions
         (\ref{m4}) and (\ref{m6}). The only difference is that $W'$ in
         (\ref{m5}) should be replaced by
\begin{equation}
               W = (1/2) \,k \,T \, \eta{(T)}                \label{m7}
\end{equation}
         where $\;W$ is the average kinetic energy of electrons per
         one degree of freedom, $\;k$ is Boltzmann's constant, $\;T$
         denotes the temperature of electrons in kelvins, and $\,\eta\,$
         is a dimensionless coefficient equal to one for particles having
         the classical distribution of velocities. Some values of $L$ and
         $\eta$ for copper, silver, and a deuteron plasma are given in
         Appendix.

            Suppose that the values of $\Phi_0$ and $u$ or $W$ for
         electron carriers in cubes $P^{(i)}$ are similar to those for
         carriers in copper at 300 K. Using the data from formulas
         (\ref{A.01}) in Appendix, we get then from (\ref{m4}) and
         (\ref{m6}) the numerical values of ${\bf E}_B$, ${\bf H}_B$
         mentioned in Section~2 for copper cube. Therefore, we obtain
\begin{equation}
            |{\bf G}_B^{(2)}| \gg 0.                    \label{m8}
\end{equation}
\vspace{0\xa}

\noindent
         {\bf{3.4. Conclusions}}\nopagebreak
\vspace{5\xa}

            From (\ref{1.1}), (\ref{m1}), (\ref{m2}), and (\ref{m3}), we
         get
\begin{eqnarray}
            {\bf G}^{(1)} =  {\bf G}_T,
       & &  {\bf G}^{(2)} =  {\bf G}_T + {\bf G}_B^{(2)}.
         \label{m9}
\end{eqnarray}
         Hence, taking into account (\ref{m8}), we obtain that at
         least one of the following inequalities must be true:
\begin{eqnarray}
            |{\bf G}^{(1)}| \gg 0,
       & &  |{\bf G}^{(2)}| \gg 0,
         \label{m10}
\end{eqnarray}
         which proves the formula (\ref{1.3}).

            As it has been already pointed out in Section 2, the result
         (\ref{m10}) does not agree with the results which can be
         obtained by means of the presently used methods of macroscopic
         averaging. These methods lead in our case to the conclusion that
         ${\bf G}^{(i)} \equiv {\bf G}_0^{(i)} \equiv 0$ $\;$ ($i=1,2$),
         because in both cases the macroscopic current and charge
         densities are equal to zero. One of the reasons of this
         contradiction is the fact that when we calculate the macroscopic
         field generated by carriers from $P^{(i)}$, the contributions to
         this field from carriers leaving or entering $P^{(i)}$ are
         discontinuous in time and do not satisfy the pertinent
         microscopic Maxwell equations. This invalidates, e.g., the
         indicated in Section 1 formulas given by de Groot (1969), and
         de Groot and Suttorp (1972).
\vspace{0\xa}\\
\begin{center}
         {\bf{4. FIELD $\:{\bf{E}}_{BS}\,$ GENERATED
         BY A SURFACE $\:\partial{P}_S\:$
         \nopagebreak \\IN AN INHOMOGENEOUS CONDUCTOR }}\nopagebreak
\end{center}
\vspace{-5\xa}

            Let $\;R_0\;$, $\psi\;\;$ ($R_0 \gg \psi$) be positive
         constants. Consider a cube $P'$ defined by
\begin{equation}
              x'_i \in [-2R_0,2R_0], \;\;\;\;\;\;\;\;\;\; i=1,2,3.
         \label{p1}
\end{equation}
            The cube contains an inhomogeneous alloy of silver and
         copper. On every surface $\;x'_3=constant\;$ the alloy is
         homogeneous. For $\;x'_3 < -\psi\;$ the medium in $P'$ is a
         homogeneous copper while for $\;x'_3 > \psi\;$ the medium in
         $P'$ is a homogeneous silver. Let $\;2\psi=0.01{\rm\3cm}$,
         $\;R_0=1{\rm\3cm}$.

            Consider in $P'$ a cylinder $P_S$ defined by
\begin{equation}
         (x_1')^2 + (x_2')^2 \le
         R_0,\;\;\;\;\;\;\;|x_3'|\le1.1\psi,     \label{p2}
\end{equation}
            We denote the lateral surface of $P_S$ by
         $\;\partial_L{P_S}\;$ and the bases of $P_S$ by
         $\;\partial_{b1}{P_S}\;$ and $\;\partial_{b2}{P_S}\;$. The base
         $\;\partial_{b1}{P_S}\;$ lies on the surface $\;x'_3=-1.1\psi$.
         Assume that the boundary of $P_S$ is impenetrable for electrons.
         We shall calculate the boundary acceleration field
         $\;{\bf{E}}_{BS}\;$ generated by electrons from $P$ ($P=P'-P_S$)
         accelerated at the boundary of $P_S$. Let the observation point
         $\bf{x}$ lie in the center of $P_S$, i.e.  ${\bf{x}}=(0,0,0)$.
         Let $\;{\bf{E}}_{BS1}$, $\;{\bf{E}}_{BS2}$, $\;{\bf{E}}_{BSL}\;$
         be the contributions to $\;{\bf{E}}_{BS}\;$ generated by
         electrons accelerated at $\;\partial_{b1}P_S$,
         $\;\partial_{b2}P_S$, and $\;\partial_{L}P_S$, respectively.

            Integrating formula (\ref{m4}) with respect to $S$ over the
         base $\;\partial_{b1}P_S\;$ in suitable polar coordinates, one
         obtains:
\begin{eqnarray}
         {\bf{E}}_{BS1} = 2 \pi L {\bf{K}} R_{01}^{-1}(R_{01}-1.1\psi)^2
         & {\rm{where}} & R_{01} = (R_0^2 +1.1^2\psi^2)^{(1/2)}
         \label{pp2}
\end{eqnarray}
         (a more detailed derivation of formulas (\ref{pp2}) to
         (\ref{5.3.10}) is given in sections 5.2 and 5.3 of my previous
         pertinent paper, available electronically). In our case
         ($R_0\gg\psi$) we get with a good accuracy
         $\;{\bf{E}}_{BS1}=2\pi L_{Cu} R_0 {\bf{K}}$ where $L_{Cu}$ is
         the value of $L$ for electrons in copper.  Similarly,
         ${\bf{E}}_{BS2}=-2\pi L_{Ag} R_0 {\bf{K}}$. Hence
\begin{eqnarray}
         {\bf{E}}_{BSb} \equiv {\bf{E}}_{BS1} + {\bf{E}}_{BS2} = L_2 R_0
         {\bf{K}}, && L_2 = 2 \pi (L_{Cu} - L_{Ag}).
         \label{p3}
\end{eqnarray}
            If $\kappa$ is sufficiently small, then it can be easily
         proved that the field ${\;\bf{E}}_{BSL}\;$ can be estimated by
         integrating formula (\ref{m4}) with suitably changed $\bf{K}$
         over $\;\partial_{L}P_S$. In our case ($2\psi=0.01{\rm\3cm}$,
         $\;R_0=1{\rm\3cm}$) one can obtain an estimate
         $\,|{\bf{E}}_{BSL}|/|{\bf{E}}_{BSb}|<3\!\times\!10^{-4}$ if one
         puts $\Phi_0$ in the integrand twice as big as $\Phi_0$ for
         copper. Therefore, with a good accuracy
\begin{eqnarray}
         & {\bf{E}}_{BS} = {\bf{E}}_{BSb} + {\bf{E}}_{BSL} = L_2 R_0
         {\bf{K}}. & \label{pp3}
\end{eqnarray}
            From (\ref{p3}) and (\ref{A.01}) we get $\,L_2=-
         3.9\!\times\!10^{15}\3 {\rm{V}\3 {m}^{-2}}\,$ at $\,300{\rm\3
         K}$ and
\begin{equation}
         |{\bf{E}}_{BS}|>3.5 \!\times\!10^{13}\3{\rm{V}\3 {m}^{-1}}
       \label{5.3.10}
\end{equation}
            for $\;R_0=1{\rm\3cm\;}$. We see that the field
         $\;{\bf{E}}_{BS}\;$ generated by particles from $P'-P_S$
         accelerated in the immediate vicinity of the whole boundary
         $\;\partial{P}_S\;$ may reach surprisingly great values even
         when the diameter of $P_S$ is small.
\vspace{0\xa}\\
\begin{center}
         {\bf{5. REMARKS}}\nopagebreak
\end{center}
\vspace{-10\xa}

            This paper is a shorter and simpler version of my earlier
         paper described in Introduction. These of the more important
         results of that paper which have not been derived in the present
         paper are concisely formulated without proofs in the following
         remarks. Some methodological remarks have been also retained and
         are given in the next section. The earlier paper is available
         from the author on request electronically.
\vspace{10\xa}

            {\bf{Remark 1}}. It can be proved that formulas (\ref{m4}),
         (\ref{m6}) with expressions (\ref{m5}), where $W'$ is replaced
         by $W$ from (\ref{m7}), are also valid for a general medium
         where the thickness of the boundary layer $\kappa$ is comparable
         to, or greater than, the mean free path $l$ of carriers in $P$
         (the definition of $\kappa$ in such a general case is given in
         the earlier paper). They are also valid for curved boundaries
         when the moduli of radii of normal curvatures are sufficiently
         big in comparison with $\kappa$ and~$l$, and in some cases when
         the latter condition is not satisfied.
\vspace{10\xa}

            {\bf{Remark 2}}. If the ions in $P$ move and collide with the
         boundary $\partial_2{P}$, then we must also take into account
         the boundary acceleration field produced by ions. This field is
         described by the same formulas (\ref{m4}), (\ref{m6}) with the
         data for electrons in expressions (\ref{m5}) and (\ref{m7})
         replaced by the data for ions. Let the subscripts $i$ and $e$
         label the quantities referring, respectively, to ions or nuclei
         and electrons. If, for instance, all ions have the same charge
         $Q_i$ and mass $m_i$, we get from (\ref{m5}) and (\ref{m7})
\begin{displaymath}
         |L_i/L_e| = (\eta_i/\eta_e)\, m_e/m_i
\end{displaymath}
         as $|Q_i\Phi_{0i}|=|Q_e\Phi_{0e}|$. In this and in the other
         cases, the contribution to ${\bf E}_{B}$, ${\bf H}_{B}$ from
         nuclei or ions may be usually neglected in comparison with the
         contribution from electrons, since the ratio $m_e/m_i$ is small
         and $\eta_i/\eta_e\le 1$.
\vspace{10\xa}

            {\bf{Remark 3}}. Formulas (\ref{m4}), (\ref{m6}) with
         (\ref{m5}) and (\ref{m7}) result from expressions (\ref{2.4.5}).
         The latter are the first terms of the expansions with respect to
         the powers of $u/c$ of the exact expressions. Therefore,
         (\ref{m4}) and (\ref{m6}) are also the first terms of the
         expansions of the exact boundary acceleration field ${\bf
         G}_B^R$.  If $P$ contains a homogeneous medium, then the
         formulas (\ref{m4}), (\ref{m6}) integrated over the closed
         surface $\partial{P}$ vanish. In our cases the values of $L$ are
         very large and the values of $u/c$ are not very small (see
         Appendix). Therefore, one should consider the influence of
         further terms of the expansions of ${\bf G}_B^R$. Some partial
         results suggest that the second terms of these expansions
         integrated over $\partial{P}$ may be different from zero for
         plasma in magnetic traps, even when the plasma were homogeneous.
\vspace{10\xa}

            {\bf{Remark 4}}. Let ${\bf G}_0\not=0$. Assume that ${\bf
         G}_0= {\bf G}_{01}+{\bf G}_{02}$ where ${\bf G}_{01}$ does not
         depend on $t$ and $|{\bf G}_{02}|$ is sufficiently small. In
         general, the density ${\bf J}$ and the field ${\bf G}_{01}$ may
         influence the velocity distribution functions of particles in $P$.
         Hence the fields ${\bf G}_B$ and ${\bf G}_T$ may depend on ${\bf
         J}$ and $\rho$. However, if we assume that (\ref{1.2}) holds for
         ${\bf J}\not=0$, $\rho\not=0$ when $\partial_2P$=$\partial P\/$
         or that $|{\bf J}|$, $|\rho|$ are sufficiently small, then
         formulas (\ref{1.8}) and (\ref{1.9}) remain valid.
\vspace{0\xa}\\
\begin{center}
         {\bf{6. METHODOLOGICAL REMARKS}}\nopagebreak
\end{center}
\vspace{-10\xa}

            {\bf{Remark M1}}. The aim of this paper is a presentation of
         some new results deduced from the Maxwell equations with
         microscopic charge and current densities. These results concern
         averaged macroscopic electromagnetic field generated by large
         sets of moving charged particles. We do not know the position
         and velocity of any particular particle, and may use only
         statistical information given by velocity distribution functions
         for various particle species in $P$. Nevertheless, we assume
         that the positions and velocities of particles are continuous
         functions of time. Such an assumption is necessary to use, e.g.,
         the Li\'{e}nard-Wiechert formulas. Without it any results
         concerning the electromagnetic field generated by sets of moving
         carriers are not a deduction from the Maxwell equations with
         microscopic sources even if they are an inference from some
         inductive extensions of these equations used in quantum
         theories. Therefore, we shall introduce a natural model based on
         this assumption and taking into account quantum considerations.

            Let the region $P$ contain $N$ carriers with the same total
         electric charge $Q$ and rest mass $m$, and let $b$ denote a
         sufficiently small positive number. Let the macroscopic number
         density of these carriers be $\Phi({\bf x},t)$ and their
         velocity distribution function be $f({\bf x,u},t)$. For sets of
         degenerate particles, such as electron plasma in metal
         conductors, the values of $f({\bf x,u},t)$ may result from
         quantum theories. We define in $P$ a set of $N$ privileged
         points with the following properties. The positions and
         velocities of these points are continuous functions of time, and
         their accelerations are piecewise continuous functions. Their
         trajectories do not cross $\partial_2 P$. At a time $t_0$ their
         macroscopic number density is equal to $\Phi({\bf x},t_0)$. If
         the distance between two privileged points is less than $2b$,
         then they are strongly accelerated in random directions.
         Suitably choosing the distribution of such accelerations, we may
         ensure that the number density and velocity distribution of
         privileged points in $P$ are given by $\Phi({\bf x},t)$ and
         $f({\bf x,u},t)$, respectively.

            We regard each privileged point as a point belonging to
         the region occupied by a particle with the charge $Q$ and rest
         mass $m$. When we write that particle $p$ is at a point $\bf x$
         at a time $t$, we mean that privileged point $p$ is at
         $\bf x$ at $t$. All species of particles in $P$ can be
         represented in this way. We may allow interactions between
         privileged points of different kinds of particles if we care
         about a more realistic description of collisions.

             Procedures equivalent to the setting of privileged points are
         universally used in statistical physics. For instance, they were
         explicitly or implicitly used by de~Groot and Suttorp (1972),
         Ferziger and Kaper (1972), Landau and Lifshitz (1964), Schram
         (1991).
\vspace{10\xa}

            {\bf{Remark M2}}. Let $\rd^3x$ be a small volume in the
         system $X$ in $P$, $u_n$ be the components of $\bf{u}$ in $X$,
         and $\rd^3u$ be a small volume in the velocity space $u_n$.
         Assume that at a time $t$ the expected number $\rd{N}({\bf
         x,u},t)$ of particles of a given species in the volume element
         $\rd^3x$ located at {\bf x}, whose velocities lie in $\rd^3u$
         about velocity {\bf u} is given with a sufficient accuracy by
         the formula
\begin{displaymath}
         \rd{N}({\bf x,u},t) = f({\bf x,u})\, \rd^3x\, \rd^3u
         \label{2.2.1}
\end{displaymath}
         where the velocity distribution function $f({\bf x,u})$ does
         not depend on $t$.

            Taking into account regions in the vicinity of the boundary
         $\partial_2 P$, it may be sometimes convenient to consider
         subregions of $P\/$ which are too small to contain a large
         number of particles (see, e.g., the next remark). We may then
         take into account a set of identical regions $P_n$ ($n\,$=$\,
         $1,2,\dots,$I_{en}$). In each region $P_m$ ($m\,$$>$1) the
         privileged points of all particles are located at a time $t$ in
         the same points in which they were located in $P_{m-1}$ at a
         time $t\,$-$\,\delta{t}$.  Considering all privileged points
         from all $P_m$, we obtain for them the velocity distribution
         function $f_{en}$ equal to $I_{en}f({\bf x,u})$. The reasoning
         may be then carried out for $f_{en}$ instead of $f$. The final
         result should be divided by $I_{en}$. This approach makes
         $\rd{N}$ well defined even when $\rd{N}<1$. This is one of
         possible procedures making the concept of the velocity
         distribution function useful also in idealized or averaged
         microscopic regions. Such an averaging may be performed, e.g.,
         in the microscopic vicinity of $\partial_2 P$ on microscopic
         layers parallel to $\partial_2 P$.
\vspace{10\xa}

            {\bf{Remark M3}}. We assume throughout this paper that the
         boundary $\partial_2P$ is impenetrable for carriers of electric
         charge.  The form of the functions $f({\bf x,u})$ in a vicinity
         of $\partial_2P$ depends on the averaged distribution of forces
         acting on carriers in the neighborhood of $\partial_2 P$. This
         distribution depends on the kind of medium in $P$ and its
         theoretical idealization. We can also introduce artificial
         boundaries and postulate extremal or not typical mechanisms of
         their impenetrability. In each case the forces acting in the
         vicinity of an impenetrable surface cause the averaged
         acceleration of particles to have a not vanishing resultant
         pointing into $P$. Consequently the number densities of carriers
         and the functions $f({\bf x,u})$ for various species of
         particles must, in general, depend on the distance from the
         point ${\bf x}$ to such a surface for ${\bf x}$ in a
         neighborhood of this surface. The only exception concerns the
         case when the barrier of forces making the boundary
         $\partial_2{P}$ impenetrable is sufficiently strong and steep to
         decelerate and accelerate the carriers within a distance which
         is very small in comparison with the average distance between
         particles in $P$.
\vspace{10\xa}

            {\bf{Remark M4}}. Consider a moving charged particle with the
         total electric charge $Q$. The electromagnetic field
         $\bf\cal{E}$, $\bf\cal{H}$ generated by this particle and
         observed at the point ${\bf x}$ and time $t$ may be written as
\begin{displaymath}
   {\bf{\cal E}(x},t) = {\bf e(x},t)+{\bf e'(x},t), \;\;\;\;\;
   {\bf{\cal H}(x},t) = {\bf h(x},t)+{\bf h'(x},t), \label{2.4.1a}
\end{displaymath}
         where the field $\bf e$, $\bf h$ is equal to the field which
         would be generated in a vacuum by the charge $Q$ located in the
         privileged point related to this particle. The field $\bf e'$,
         $\bf h'$ is a correction to the field $\bf e$, $\bf h$. This
         correction may be produced owing to the following reasons: the
         charge is not located in the privileged point, its density is
         finite, the charge density is strongly inhomogeneous and of
         opposite sign in some subregions of the particle, there are
         electric currents flowing and/or oscillating inside the
         particle, there are electric currents caused by the rotation of
         the whole particle. Let ${\bf\cal
         {E'}}_{\!\!\!\scriptscriptstyle\Sigma}({\bf x},t)$, ${\bf\cal
         {H'}}_{\!\!\scriptscriptstyle\Sigma}({\bf x},t)$ be the sum of
         corrections $\bf e'$, $\bf h'$ from all particles in $P$,
         averaged in a short interval of time. The value of this sum
         depends on the kind of medium in $P$. In this paper we assume
         that if $\bf{x}$ is outside $P$ and its immediate vicinity, then
         ${\bf\cal{E'}}_{\!\!\!\scriptscriptstyle\Sigma}({\bf x},t)=0$,
         ${\bf\cal H'}_{\!\!\scriptscriptstyle\Sigma}({\bf x},t)=0$ with
         a sufficient accuracy. This is equivalent to the assumption that
         outside $P$ and its immediate vicinity we may neglect the total
         field of all electric and magnetic multipoles which may be
         introduced in order to describe the fields $\bf e'$, $\bf h'$
         generated by all particles in $P$.
\vspace{10\xa}

            {\bf{Remark M5}}. The field ${\bf G}_0$ generated by ${\bf
         {J}}$ and $\rho$ in $P$ is usually calculated by means of the
         retarded vector and scalar potentials ${\bf{A}}$ and $\varphi$.
         If, for instance, the medium in and outside $P$ is homogeneous,
         and its permeability and dielectric constant are equal to 1,
         then the potential ${\bf{A}}$ is given by
\begin{displaymath}
 {\bf A(x},t) = \frac{\mu_0}{4\pi}\!\int_P\! d^3x'\,|{\bf x-x'}|^{-1}
         \,{\bf J(x}',t_d)
\end{displaymath}
         where $t_d=t-c^{-1}|{\bf x-x'}|$, and the potential
         $\varphi({\bf x},t)$ is given by the analogical expression (see,
         e.g., Bochenek 1961). These potentials may be used to calculate
         the field ${\bf{G}}_0$ if the Lorentz condition ${\bf
         \nabla\cdot{A}}+c^{-2}\:{\partial\varphi}/{\partial t}=0$ is
         satisfied. This always takes place when the normal to
         $\partial{P}$ component of ${\bf{J}}$ vanish identically on
         $\partial{P}$. It is a consequence of results given, e.g., by
         Stratton (1941) or Jones (1964). If this last assumption is not
         satisfied, then the field ${\bf{G}}_0$ is not defined while
         ${\bf{G}}$ remains well defined.
\vspace{0\xa}\\

\noindent
         {\bf{Acknowledgements}} \nopagebreak
\vspace{0\xa}

            The author acknowledges the helpful attitude of Professors
         Micha{\l} Kleiber, Adam Ciarkowski, and Marek Matczy\'{n}ski
         from the Institute of Fundamental Technological Research
         concerning the organization of his workplace.
\vspace{0\xa}\\
\renewcommand{\theequation}{A\arabic{equation}}
\begin{center}
         {\bf{APPENDIX}} \nopagebreak
\end{center}
\setcounter{equation}{0}
\vspace{-10\xa}

            We shall calculate here the numerical values of $L$ defined
         in (\ref{m5}) --- with $W'$ replaced by $W$ given in (\ref{m7})
         --- for electron carriers in copper and silver at a temperature
         of $\,300{\rm\3 K}$ and for electrons in a deuteron plasma
         prepared for a controlled thermonuclear fusion reaction, with
         the number density of nuclei $\Phi_{0D}=10^{22}\3{\rm{m}^{-3}}$,
         the temperature of nuclei of $4\!\times\!10^8\3\rm{K}$, and the
         temperature of electrons $T=10^7\3\rm{K}$. The quantities
         referring to electrons in copper and silver will be denoted by
         subscripts Cu and Ag, respectively, while those referring to
         electrons in the deuteron plasma will have subscript D.

            The following constants are used: $c =
         3.00\!\times\!10^8{\rm\3 m\3 s^{-1}}$, $c\varepsilon_o=
         2.65\!\times\!10^{-3}\rm\3 F\3s^{-1}$, $h_P=
         6.63\!\times\!10^{-34}\rm\3{J}\3{s}$ (Planck's constant), $k =
         1,38\!\times\!10^{-23}\rm\3 J\3K^{-1}$ (Boltzmann's constant),
         $m_e= 9.11\!\times\!10^{-31}\rm\3 kg$ (the rest mass of
         electron), $N_A=6,02\!\times\!10^{23}\rm\3 mol^{-1}$ (Avogadro's
         number), $Q = -1.60\!\times\!10^{-19}\rm\3 C$ (the electric
         charge of electron).

            We define
\begin{equation}
             W_{av} = 3 W = (3/2)\,kT \eta,       \label{A.1}
\end{equation}
\begin{equation}
               (1/2)\,m_e u_{av}^2 = W_{av}.       \label{A.2}
\end{equation}
            We see that $u_{av}$ is an averaged speed of electrons and
         $W_{av}$ is their average kinetic energy in the non-relativistic
         range of velocities when these electrons move in all 3 space
         directions. It will be shown that for the electron plasma in
         copper and silver at 300\rm\3{K} we get
         $c^{-1}u_{av}<5\!\times\!10^{-3}$. Therefore we may neglect the
         relativistic corrections to the mass and kinetic energy of
         electrons. As far as electrons in the plasma at $10^7{\rm\3 K}$
         are concerned, we receive $c^{-1}u_{av}<0.08$. In this case we
         also neglect the relativistic corrections, since this may cause
         at most a few per cent error of $L_D$.

            Assume that the Fermi-Dirac statistics correctly describes
         the energy and velocity distributions of electrons in electron
         plasma. If we consider electron plasma contained in a cube with
         a volume $v$ in physical space, then the volume of the
         elementary cell in the momentum space is $h_P^3/v$. Each cell
         contains at most 2 electrons. Hence, if $\zeta_0$ denotes the
         maximum energy of electrons at 0\3{K}, and $W_{0av}$ is the
         average energy corresponding to $\zeta_0$, we get
\begin{eqnarray}
        \zeta_0 =h_P^2(8m_e)^{-1}({3\Phi_0} /{\pi})^{2/3},
         & &   W_{0av} = 0.6\,\zeta_0.
     \label{A.4}
\end{eqnarray}
         If $kT/\zeta_0\ll 1$, then we may neglect the increase of the
         average energy caused by temperature and put
\begin{equation}
         W_{av} = W_{0av}.               \label{A.5}
\end{equation}
         If $kT/\zeta_0\gg 1$, then the Fermi-Dirac statistics becomes
         that of Maxwell and we may set $\eta=1$. The exact analysis
         (see, e.g., Weizel 1955) gives for $kT/\zeta_0$ much smaller
         than 1:
\begin{displaymath}
        W_{av}=W_{0av}[1+({5\pi^2}/{12})(kT/\zeta_0)^2] +O[(kT/\zeta_0)^4].
      \label{A.6}
\end{displaymath}
         If $kT/\zeta_0$ is much greater than 1 then from formulas in
         Weizel's book one may infer (see also, e.g., Tamm 1957) that
         $\eta=1$ with a good accuracy if
\begin{eqnarray*}
          \Phi_0\, h_P^3(2\pi m_e kT)^{-3/2}<0.01 &\mbox{ or }&
         T>1.3\!\times\!10^{-13}\,\Phi_0^{2/3}.   \label{A.7}
\end{eqnarray*}
         Therefore, for $\Phi_{0D}=10^{22}{\rm\3 m}^{-3}$ and
         $T=10^7{\rm\3 K}$, we get $\eta_D=1$. From (\ref{A.1}) and
         (\ref{A.2}) we receive then $c^{-1}u_{avD}<0.08$.

            We calculate now $\zeta_0$ for copper and silver. The mass of
         1 mole of Cu is $63.5{\rm\3 g}$ and of Ag is $108{\rm\3 g}$.
         Assume that the specific mass at 300\3{K} of Cu is $8.9\rm\3
         Mg\3 m^{-3}$, and of Ag is $10.5\rm\3 Mg\3 m^{-3}$.  Hence the
         number densities of atoms of Cu and Ag are, respectively,
         $8.4\!\times\!10^{28}\rm\3 m^{-3}$ and $5.9\times 10^{28}\rm\3
         m^{-3}$. The average number of electrons given to the electron
         plasma by one atom at 300\3{K} is, respectively, 1.3 and 1.03
         (see, e.g., Szczeniowski 1955, Clemmow 1973).  Therefore, at
         300\3{K}
\begin{equation}
        \!\!\!\!\!\!\!\!\Phi_{0Cu}=11\!\times\!10^{28}\3{\rm m}^{-3},\;\;
        \;\;\Phi_{0Ag}=6.1\!\times\!10^{28}\3{\rm m}^{-3},  \label{A.8}
\end{equation}
\vspace{5\xa}
\begin{equation}
\begin{array}{lc}
         \zeta_{0Cu}=1.3\!\times\!10^{-18}\3{\rm J},
         & W_{0avCu}=0.78\!\times\!10^{-18}\3{\rm J},\\ \\
         \zeta_{0Ag}=0.88\!\times\!10^{-18}\3{\rm J},
         & W_{0avAg}=0.53\!\times\!10^{-18}\3{\rm J}.
\end{array}
         \label{A.9}
\end{equation}
\vspace{5\xa}
         Formulas (\ref{A.9}) specify the maximum and average energies
         which the electrons would have at 0\3{K}, were the numbers
         densities of electrons at 0\3{K} equal to $\Phi_{0Cu}$ and
         $\Phi_{0Ag}$ given by (\ref{A.8}). For T=300\3{K} we have
         $\;kT/\zeta_{0Cu}=3.2\!\times\!10^{-3}$, $\;\;kT/\zeta_{0Ag}
         =4.7\!\times\!10^{-3}$. Therefore we may use (\ref{A.5}). From
         the approximate formula $\;u_{av}=(2W_{0av} m_e^{-1})^{1/2}\;$
         we get $\;c^{-1}u_{avCu}<4.6\!\times\!10^{-3}$, $\;\;c^{-1}
         u_{avAg}<3.9\!\times\!10^{-3}$. It results from the Fermi-Dirac
         statistics that for nearly all electrons in copper and silver
         $\;c^{-1}u<7\!\times\!10^{-3}\;$ at $\;T=300\rm\3 K$.

            Taking into account (\ref{A.1}), (\ref{A.5}), (\ref{A.8}),
         (\ref{A.9}), we obtain from (\ref{m5}) and (\ref{m7}) that at
         $300\rm\3 K$
\begin{equation}
\begin{array}{cc}
         L_{Cu} = - 1.0 \!\times\!10^{15}\3{\rm{V}\3 {m}^{-2}},
         & c\,\varepsilon_0\, L_{Cu} = - 2.7 \!\times\!10^{12}\3{\rm{A}\3
         {m}^{-2}},\\ \\
         L_{Ag} = - 3.8 \!\times\!10^{14}\3{\rm{V}\3 {m}^{-2}},
         & c\,\varepsilon_0\, L_{Ag} = - 1.0 \!\times\!10^{12}\3{\rm{A}\3
         {m}^{-2}}.
\end{array}
         \label{A.01}
\end{equation}
         On grounds of (\ref{A.1}), (\ref{A.4}), and (\ref{A.5}) we
         notice that $L_{Cu}$ and $L_{Ag}$ should not practically depend
         on temperature in an interval within which the number densities
         (\ref{A.8}) remain nearly constant. Replacing $W_{av}$ in
         (\ref{A.1}) by $W_{0av}$ from (\ref{A.9}), we receive for
         $T=300\rm\3 K$
\begin{equation}
       \eta_{Cu}=126,\;\;\; \eta_{Ag}= 85.          \label{A.11}
\end{equation}
         Taking into account $\eta_D=1$, we obtain for $T=10^7{\rm\3 K}$
\begin{eqnarray}
         L_D = - 2.4 \!\times\!10^{10}\3{\rm{V}\3{m}^{-2}}, &
         & c\,\varepsilon_0\, L_D = - 6 \!\times\!10^{7}\3{\rm{A}\3{m}^{-2}}.
         \label{A.02}
\end{eqnarray}
\noindent
\vspace{10\xa}\\
         {\bf{References}}
\setlength{\parindent}{0in}
\nopagebreak
\begin{description}
\vspace{-10\xa}

\item Bochenek, K., {\em Metody analizy p\'{o}l
         elektromagnetycznych}. (Pa\'{n}stwowe Wydawnictwo Naukowe,
         Warszawa--Wroc{\l}aw, 1961), pp. 30--31.
\vspace{-15\xa}

\item Clemmow, P. C. and Dougherty, J. P.,
         {\em Electrodynamics of particles and plasmas}.
         (Addison-Wesley, Reading, Ma., 1969), pp. 37--38.
\vspace{-15\xa}

\item Clemmow, P. C., {\em An introduction to electromagnetic
         theory}. (Cambridge University Press, 1973), p. 257.
\vspace{-15\xa}

\item de Groot, S. R., {\em The Maxwell equations}.
         Studies in Statistical Mechanics (North-Holland, Amsterdam,
         1969), pp. 11-37.
\vspace{-15\xa}

\item de Groot, S. R. and Suttorp, L. G., {\em Foundations
         of electrodynamics}. (North-Holland, Amsterdam, 1972), pp.
         3--307.
\vspace{-15\xa}

\item Ferziger J. H. and Kaper, H. G., {\em Mathematical
         theory of transport processes in gases}. (North-Holland,
         Amsterdam, 1972).
\vspace{-15\xa}

\item Jones, D. S., {\em The theory of electromagnetism}.
         (Pergamon Press, Oxford, 1964), p. 40.
\vspace{-15\xa}

\item Landau L. D. and Lifshitz, E. M., {\em Statisti\v{c}eskaja
         fizika}, 2nd edn. (Nauka, Moskva, 1964).
\vspace{-15\xa}

\item Li\'{e}nard, A., "Champ \'{e}lectrique et
         magn\'{e}tique produit par une charge \'{e}lectrique
         concentr\'{e}e en un point et anim\'{e}e d'un mouvement
         quelconque," {L'\'{e}clairage \'{e}lectrique} {\bf 16}, 5,
         53, 106 (1898).
\vspace{-15\xa}

\item Lorentz, H. A., {Proc. Roy. Acad. Amsterdam} 254 (1902). \vspace{-15\xa}

\item Lorentz, H. A., "Weiterbildung der Maxwellschen Theorie:
         Elektronentheorie," {Enc. Math. Wiss.} {\bf 14} (1904).
\vspace{-15\xa}

\item Lorentz, H. A., {\em The theory of electrons}, 2nd
         edn. (Dover, New York, 1915).
\vspace{-15\xa}

\item Schram, P. P. J. M., {\em Kinetic theory of gases
         and plasmas}. (Kluwer Academic Publishers, Dordrecht, 1991).
\vspace{-15\xa}

\item Stratton, J. A., {\em Electromagnetic theory}.
         (McGraw-Hill, New York and London, 1941), pp. 429--430.
\vspace{-15\xa}

\item Sza{\l}ek, M. A., "Pauli versus the Maxwell equations
         and the Biot-Savart law," {Phys. Essays} {\bf10},
         95 (1997).
\vspace{-15\xa}

\item Szczeniowski, S., {\em Fizyka do\'{s}wiadczalna,
         cz\c{e}\'{s}\'{c} III, elektryczno\'{s}\'{c} i magnetyzm}.
         (Pa\'{n}stwowe Wydawnictwo Naukowe, Warszawa, 1955), p.~217.
\vspace{-15\xa}

\item Tamm, I. E., {\em Osnovy teorii elektri\v{c}estva},
         7th edn. (Gosudarstvennye Izdatelstvo
         Tehniko$\:$--Teoreti\v{c}eskoj Literatury, Moskva, 1957),
         \S 41, p.198.
\vspace{-15\xa}

\item Weizel, W., {\em Lehrbuch der theoretischen Physik}.
         (Springer 1955), chapter L, section~IV,~\S1.
\vspace{-15\xa}

\item Wiechert, E., "Elektrodynamische Elementargesetze,"
         {Archives n\'{e}erlandaises} {\bf 5}, 549 (1900).

\end{description}
\label{lastpage}

\end{document}